\documentclass[onecolumn,pra,aps,showpacs]{revtex4}

\usepackage{amssymb}
\usepackage{graphicx}
\usepackage{amsmath}
\usepackage{dcolumn}
\usepackage{bm}
\usepackage[usenames]{color}
\usepackage[normalem]{ulem}
\usepackage{comment}

\newcommand{\beq}{\begin{equation}}
\newcommand{\eeq}{\end{equation}}
\newcommand{\beqa}{\begin{eqnarray}}
\newcommand{\eeqa}{\end{eqnarray}}

\newcommand{\la}{\langle}
\newcommand{\ra}{\rangle}

\def\ajp#1{{ Am.\ J.\ Phys.} {\bf #1}}

\def\nat#1{{ Nature} {\bf#1}}

\def\nphys#1{{Nature\ Phys.} {\bf#1}}

\def\ol#1{{ Opt.\ Lett.} {\bf#1}}
\def\pla#1{{ Phys.\ Lett. A\/} {\bf#1}}

\def\pr#1{{ Phys.\ Rev. } {\bf#1}}
\def\pra#1{{ Phys.\ Rev. A\/} {\bf#1}}

\def\prl#1{{ Phys.\ Rev.\ Lett.} {\bf#1}}

\newcommand\st{\textcolor{red}{\bgroup}\markoverwith
{\textcolor{red}{\rule[.5ex]{2pt}{0.4pt}}}\ULon}
\begin{document}

\title{Bell Violation for Unknown Continuous-Variable States}
\author{X.-F. Qian}
\email{xfqian@pas.rochester.edu}
\author{C.J. Broadbent}
\author{J.H. Eberly}
\affiliation{Rochester Theory Center and the Department of Physics
\&
Astronomy\\
University of Rochester, Rochester, New York 14627}
\date{\today }

\begin{abstract}
We describe a new Bell test for two-particle entangled systems that
engages an unbounded continuous variable. The continuous variable
state is allowed to be arbitrary and inaccessible to direct
measurements. A systematic method is introduced to perform the
required measurements indirectly. Our results provide new
perspectives on both the study of local realistic theory for
continuous-variable systems and on the nonlocal control theory of
quantum information.
\end{abstract}


\maketitle

The issue of incompatibility between local realism and the
completeness of quantum mechanics was originally raised for
unbounded continuous variables in two-party systems by Einstein,
Podolsky, and Rosen (EPR) \cite{EPR35}. Experiments to test local
realism based on inequalities proposed by Bell \cite{Bell64} and his
followers \cite{CHSH69} imply, as is well known, that classical
realism must be discarded as the basis for a universal theory. This
has been repeatedly demonstrated in experiments with discrete
variable systems \cite{Clauser-Freedman, Fry-Thompson,
Aspect-etal82, Rowe-etal01, Ansmann-etal09}.

Methods for testing local realism in continuous-variable systems
have been proposed in order to advance the goal of reaching a
completely loophole-free conclusion, and experimental tests on
continuous-variable systems have been carried out
\cite{Rarity1990,Kwiat1993,Strekalov1996,Kuzmich2000,Yarnall-etal07}.
However, these tests and all continuous-variable proposals to date
\cite{Rarity1990,Kwiat1993,Strekalov1996,Kuzmich2000,Yarnall-etal07,Franson1989,Banaszek,Gilchrist-etal98,
Munro-Milburn98, Chen-etal02, Wenger-etal03, Brukner2003} fall short
because they rely on advance knowledge of the state under test. These methods fail whenever the state
under test is unknown because then there is no basis by which
measurement strategies can be guaranteed effective. One reason is
that non-local correlations present in the original state can evade
detection under dimensional reduction \cite{Broadbent-etal}, as may
happen, for example, in pursuing pseudo-spin
\cite{Chen-etal02,Brukner2003,Yarnall-etal07} or binning
\cite{Wenger-etal03} methods.  An exceptiona l approach by E.G. Cavalcanti et al \cite{CFRD} leads to a
continuous multipartite inequality that doesn't rely on advance
knowledge of the state under test. However, to construct their
inequality, operator commutation relations must be ignored, which
also eliminates a large category of local realistic theories from
test -- see Q. Sun et al \cite{CFRD}. Additionally, violation of these inequalities may
not be possible with only two parties -- see A. Salles et al \cite{CFRD}.

Thus two obstacles that have not yet been overcome are these: to
derive a standard Bell-CHSH inequality \cite{CHSH69} for an
arbitrary and unknown bipartite input state in an unbounded
continuous-variable state space, and to describe a currently
feasible experimental method for its test. There are significant
fundamental and practical reasons for solving this problem. On the
fundamental side, a clear understanding of the domains of
continuous-variable space which are incompatible with local realism
remains to be achieved. More practically, in recent years
paradigm-shifting quantum technologies have been developed which
depend upon Bell non-locality in theory, and in some cases require
the experimental violation of a Bell inequality of an unknown state
\cite{Brunner2013}. Methods which permit Bell-CHSH inequalities to
be formed and then tested on unknown states in continuous-variable
systems may aid in the development and implementation of these
technologies.

In this Letter we take a significant step in the way to overcoming
both obstacles. To provide easy visualization, we address both
issues in a specific scenario using the following two-photon
down-conversion state:
\begin{align} \label{psiAB}
|\psi_{AB} \ra = &\cos\theta\, |H\ra_A \otimes\int d\vec q ~h(\vec
q)
|\vec q\ra_B  \nonumber \\
&+ \sin\theta\,|V\ra_A \otimes \int d\vec q ~v(\vec q) |\vec q\ra_B,
\end{align}
where $|\vec{q}\ra_B$ is one of a continuum of delta-normalized
one-photon transverse momentum states of photon $B$, and $|H\ra_A$
and $|V\ra_A$ denote horizontally and vertically polarized quantum
states of photon $A$. We assume that the transverse momentum state of photon $A$
and the polarization of photon $B$ factor out of the quantum state,
and therefore need not be indicated.

The $\sin\theta$ and $\cos\theta$ factors are included in writing
$|\psi_{AB}\ra$ to preserve its unit normalization, as the complex
continuum amplitudes  $h(\vec q)$ and $v(\vec q)$ are assumed to be
unit-normalized, i.e., $\int d\vec q ~|h(\vec q)|^2 = \int d\vec q
~|v(\vec q)|^2 = 1$. Beyond normalization, nothing else is assumed
about $h(\vec q)$ and $v(\vec q)$, including the value of their
generally non-zero scalar product, \beq \label{z} \int d\vec q
~h^*(\vec q)v(\vec q) \equiv z \ne 0. \eeq

The two-photon state in (\ref{psiAB}) has an important freedom in
the amplitude functions $h(\vec q)$ and $v(\vec q)$, which are
arbitrary superpositions of the modes in continuous $\vec q$ space.
In the following we will use the term \textit{bundle} to refer to an
arbitrary superposition of $|\vec q\ra$ states. Note that this means
that it is impossible to fully determine the state (infinitely many
measurements would be required). This point is crucial because it is
the stopping point for attempts up to the present time to fully
engage a continuous degree of freedom in Bell inequality analysis.
We have overcome this roadblock, as we describe below.

It is natural to use the Schmidt analysis in considering two-party
pure state entanglement, whether discrete or continuous. The Schmidt
decomposition \cite{Schmidt} reformulates the state (\ref{psiAB}) as
\beq \label{genSchmidt} |\psi_{AB}\ra = \sum_{n=1}^\infty{\kappa
}_{n}|{u}_{n}\ra_A |f_{n}\ra_B , \eeq where the sets $\{|u_n\ra_A\}$
and $\{|f_n\ra_B\}$ are superpositions of $A$'s polarization states
and $B$'s momentum states respectively, and are derivable as the
eigenvectors of $A$'s discrete and $B$'s continuous reduced density
matrices. The $\kappa_n^2$ are the associated eigenvalues, which are
always the same for the two reduced density matrices.

Note that since party $A$ has only two dimensions it has only two
eigenvalues, and this forces all but two of $B$'s infinitely many
Schmidt eigenvalues to vanish. Thus the infinite $n$ sum in
(\ref{genSchmidt}) has only two non-zero terms, which we write: \beq
\label{Schmidt} |\psi\ra = {\kappa }_{1}|{u}_{1}\ra |f_{1}\ra
+{\kappa }_{2}|{u}_{2}\ra |f_{2}\ra, \eeq where we have dropped the
$A$ and $B$ labels because it will be easy to remember that the
discrete states belong to photon $A$ and the continuous states to
photon $B$. Here $|{u}_{1}\ra$ and $|{u}_{2}\ra$ are merely
rotations of the original polarization states $|H\ra$ and $|V\ra$,
and $|f_{1}\ra$ and $|f_{2}\ra$ are unknown bundles of $B$'s
momentum states $|\vec q\ra$, and we write them as $|f_{1}\ra = \int
d\vec q ~\varphi_1(\vec q)|\vec q\ra$ and $|f_{2}\ra = \int d\vec q
~\varphi_2(\vec q)|\vec q\ra$, with a key orthogonality property:
\beq \la f_1|f_2\ra = \int d\vec q ~\varphi_1^*(\vec
q)\varphi_2(\vec q) = 0,\label{ortho} \eeq guaranteed by the Schmidt
rearrangement \cite{Qian-Eberly11}. ${\kappa}_{1}$ and
${\kappa}_{2}$ are real positive coefficients analogous to the
$\sin\theta$ and $\cos\theta$ in (\ref{psiAB}), with
$\kappa_{1}^{2}+\kappa_{2}^{2}=1$. We note that because $h(\vec q)$
and $v(\vec q)$ are unknown, then $\{\kappa_1,\kappa_2\}$ are also
unknown. Lastly, for simplicity in the following derivation, we
assume that $z$ is real-valued which ensures that $|u_i\ra$ is
linearly polarized.

The Schmidt theorem provides an optimum result in three ways. First,
as partners for the rotated polarization states it makes two bundles
of momentum states from the (presumed unknown) amplitudes $h(\vec
q)$ and $v(\vec q)$. Second, it guarantees that those state bundles
are orthogonal, and so we have a pair of orthonormality relations
$\la{u}_{i}|{u}_{j}\ra =\la f_{i}|f_{j}\ra = \delta_{ij}$. Third,
independent of the makeup of the two bundles, the Schmidt states
$|f_1\ra$ and $|f_2\ra$ define a plane in the infinite dimensional
$|\vec q \ra$ space.

We are now much closer to Bell Inequality territory because
rotations in planes in $A$ and $B$ spaces are what the CHSH
inequality demands. But the bundles of continuum states making up
the two Schmidt states $|f_{1}\ra$ and $|f_{2}\ra$  are mysterious
because the original functions $h(\vec q)$ and $v(\vec q)$ were
unknown. There are no operators available in continuum $B$ space to
make the rotations required by the Bell-CHSH analysis. We will
describe below how to make measurements in a rotated basis in the
continuum space without rotation operators for the space, but first
let us reproduce the Bell-CHSH Inequality analysis, under the
assumption that rotations in the $|f_1\ra$-$|f_2\ra$ plane can be
controlled.

With ordinary optical components one can always undertake a rotation
of the Schmidt basis in photon $A$'s polarization space, i.e.,
\begin{eqnarray}
|{u}_{1}^{\alpha }\ra &=&\cos \alpha |{u}_{1}\ra +\sin \alpha
|{u}_{2}\ra , \\
|{u}_{2}^{\alpha }\ra &=&-\sin \alpha |{u}_{1}\ra +\cos \alpha
|{u}_{2}\ra ,
\end{eqnarray}
where $\alpha$ defines the arbitrary rotation angle. A rotated basis
$|f_{1}^{\beta }\ra$, $|f_{2}^{\beta }\ra$ of momentum space bundles
for photon $B$ can be defined similarly with $\beta$ as the rotation
angle in $\vec q$ space, while the practical matter of accomplishing
such a rotation remains temporarily an open question.

However, given these rotations, the conventional CHSH analysis of
local hidden variable theory \cite{CHSH69} can be employed. One
considers the Bell operator $\mathcal{B}$ and finds $\mathcal{B}
\leq 2$, where $\mathcal{B}$ is defined as
\begin{equation}
\mathcal{B} = C(\alpha ,\beta ) -C(\alpha ,\beta') +C(\alpha' ,\beta
) +C(\alpha',\beta').  \label{CHSH}
\end{equation}
Here $C(\alpha,\beta )$ is the CHSH correlation between photons $A$
and $B$ when the measurements are set for the angles $\alpha$ and
$\beta$, and $P_{ij}(\alpha ,\beta )$ are the joint probabilities of
finding photon $A$ in state $|{u}_{i}^{\alpha }\ra $ and photon $B$
in state $|f_{j}^{\beta}\ra $, with $i,j = 1,2$. That is,
\begin{align}
C(\alpha,\beta ) =& P_{11}(\alpha ,\beta ) - P_{12}(\alpha,\beta) \nonumber \\
&- P_{21}(\alpha,\beta) + P_{22}(\alpha,\beta).\label{correlation}
\end{align}

According to quantum mechanics, the joint probability is given as
$P_{ij}(\alpha,\beta) = \la \psi_{AB} |{u}_{i}^{\alpha }\ra
|f_{j}^{\beta }\ra \la f_{j}^{\beta }|\la {u}_{i}^{\alpha}
|\psi_{AB} \ra $, which is a joint projection in the state spaces of
both photons and has the potential to violate the CHSH inequality.
Then the Bell operator $\mathcal{B}$\ can be calculated to be
\begin{eqnarray}
\mathcal{B} &=& 2\kappa _{1}\kappa_{2}\Big[\sin 2\alpha (\sin 2\beta
-\sin 2\beta') \notag \\
&& +\sin 2\alpha'(\sin 2\beta +\sin 2\beta')\Big]  \notag \\
&& +\cos 2\alpha (\cos 2\beta -\cos 2\beta') \notag \\
&& +\cos 2\alpha'(\cos 2\beta +\cos 2\beta').
\end{eqnarray}
For the choices $\alpha=0$, $\beta =\pi/8$, $\alpha'=\alpha +\pi/4$
and $ \beta'= \beta +\pi/4$, one finds
\begin{equation}
\mathcal{B} = \sqrt{2}(2\kappa_{1}\kappa_{2} +1). \label{violate}
\end{equation}
There will be a Bell violation, $\mathcal{B} > 2$, whenever
$2\kappa_{1} \kappa_{2} > \sqrt{2}-1$. Obviously this can be
satisfied, and for the state with $\kappa_{1} = \kappa_{2} =
1/\sqrt{2}$ even the Cirelson bound is attained, i.e., $\mathcal{B}$
reaches the maximum value $2\sqrt{2}$. In fact, as was pointed by
Gisin \cite{Gisin}, the pure sate (\ref{Schmidt}) will always
violate the CHSH inequality for any non-zero $\kappa_{1}$ and $
\kappa_{2}$ if one chooses the angles $\alpha$, $\alpha'$, $\beta$
and $\beta'$ properly.

\begin{figure}[t]
\includegraphics[width=5cm]{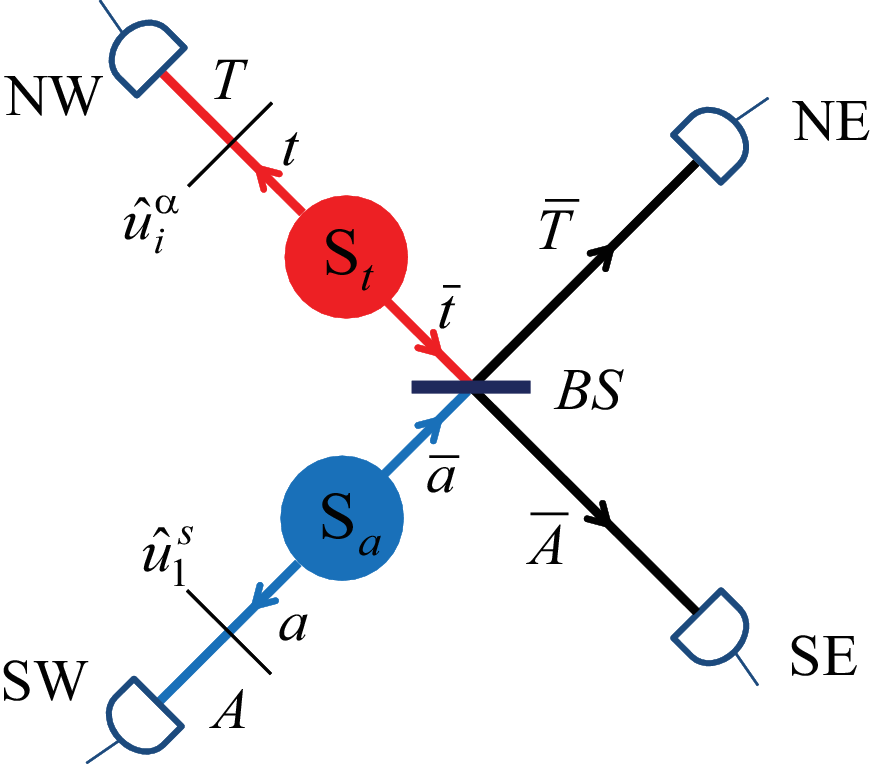}
\caption{Schematic illustration of Bell test for discrete-continuum
entangled photon pairs. The red source $S_t$ emits a photon pair
$|\psi\ra_{t\bar{t}}$, where the discrete (polarization) space of
the photon propagating towards northwest (NW) in mode $t$ is
entangled with the continuous (e.g., momentum) space of the photon
heading southeast (SE) in mode $\bar{t}$. The blue source $S_a$
emits identical photon pairs denoted as $|\psi\ra_{a\bar{a}}$ with
the discretely and continuously entangled photons propagating
towards southwest (SW) and northeast (NE) respectively. The photon
in mode $t$ passes through a polarizer $\hat{u}_i^{\alpha}$ that
passes only the polarization component $|{u}_i^{\alpha}\ra$, and
then enters mode $T$ for detection. Similar actions are taken for
the photon in mode $a$ with a polarizer $\hat{u}_1^{s}$ that passes
only the polarization component $|{u}_1^{s}\ra$. The photons in
modes $\bar{t}$ and $\bar{a}$ are combined by a 50:50 beam splitter
(BS) with two outcome modes $\bar{T}$ and $\bar{A}$ being detected.
} \label{scheme}
\end{figure}

As described above, the central hurdle to be overcome is the lack of
a method to measure the Schmidt bundles in the continuous
$|\vec{q}\ra$ space of photon $B$. As we now demonstrate, a
specially engineered auxiliary photon is sufficient to accomplish
this. The requisite auxiliary photon can be
easily created using an auxiliary entangled state which is identical
to the original state. Practical techniques for generating pairs of
identical entangled biphotons are available, as discussed in the
Supplemental Information, so we proceed with the setup sketched in
Fig.~\ref{scheme}.

Source $S_t$ emits a pair of photons in the desired
discrete-continuum entangled state, of which the Schmidt form is
\begin{equation}
|\psi \ra _{t\bar{t}} = {\kappa
}_{1}|{u}_{1}\ra_{t}|f_{1}\ra_{_{\bar{t}}} + {\kappa}_{2}
|{u}_{2}\ra_{t}|f_{2}\ra _{_{\bar{t}}}.
\end{equation}
The discretely (polarization) entangled photon in mode $t$ is
heading northwest (NW) and the continuously (momentum) entangled
photon in mode $\bar{t}$ is heading southeast (SE), illustrated by
the red paths in Fig.~\ref{scheme}. The goal of our following
analysis is to propose a Bell test, namely, measuring various
correlations in terms of joint probabilities, for such a
discrete-continuum entangled state regardless of what is known or
not known about the continuous-space photon in mode $\bar{t}$ and
whether it is accessible or not to direct measurement.

A polarization projection on basis $|{u}_{1}^{\alpha }\ra$ for the
photon in mode $t$ can be realized with a polarizer
$\hat{u}_1^{\alpha}$ that passes the $|{u}_{1}^{\alpha }\ra$
component into mode $T$, i.e., \beq |{u}_{1}^{\alpha}\ra_{t~t}\la
{u}_{1}^{\alpha}|\psi\ra_{t\bar{t}} =
|{u}_{1}^{\alpha}\ra_{T}\{\kappa_1 c_\alpha|f_1\ra_{\bar t} +
\kappa_2 s_\alpha |f_2\ra_{\bar t}\} \eeq where $c_\alpha$ and
$s_\alpha$ stand for $\cos\alpha$ and $\sin\alpha$. The probability
of this measurement outcome being realized is given by
$P_1(\alpha)=$ $_{t\bar{t}}\la \psi |{u}_{1}^{\alpha}\ra _{t~t}\la
{u}_{1}^{\alpha}|\psi \ra
_{t\bar{t}}=\kappa_1^2c_\alpha^2+\kappa_2^2s_\alpha^2$, and can be
determined experimentally by recording the number of coincidences
detected during a fixed time window in modes $(T,\bar{t})$ and
$(t,\bar{t})$ for polarizer angle $\alpha$,
\begin{align}
P_1(\alpha)=\frac{N_\alpha(T,\bar{t})}{N(t,\bar{t})},
\label{individual}
\end{align}
where $N_\alpha(T,\bar{t})$ and $N(t,\bar{t})$ are the number
of coincidences in their corresponding modes. This also gives the
value of $\kappa_1$ and $\kappa_2$ since $\kappa_1^2+\kappa_2^2=1$
as stated after \eqref{ortho}.

To determine joint probabilities, one needs to measure the continuum
space in a basis rotated by the angle $\beta$ as well, so we now
express the state in the rotated basis,
$\{|f_1^\beta\ra,|f_2^\beta\ra\}$,
\begin{align}\label{stripdemo}
|{u}_{1}^{\alpha}\ra_{t~t}\la {u}_{1}^{\alpha}|\psi\ra_{t\bar{t}}
=&|{u}_{1}^{\alpha}\ra_{T}\Big\{ (\kappa_1 c_\alpha c_\beta +
\kappa_2 s_\alpha s_\beta) |f_1^\beta\ra_{\bar t} \nonumber \\
&\quad\,+ (-\kappa_1 c_\alpha s_\beta + \kappa_2 s_\alpha c_\beta)
|f_2^\beta\ra_{\bar t}\Big\},
\end{align}
which we rewrite again as
\begin{align}
|{u}_{1}^{\alpha}\ra_{t~t}&\la {u}_{1}^{\alpha}|\psi\ra_{t\bar{t}} =
\sqrt{P_{1}(\alpha)}|{u}_{1}^{\alpha
}\ra_{T}\Big(c_{11}|f_{1}^{\beta }\ra_{\bar{t}} +
c_{12}|f_{2}^{\beta }\ra _{\bar{t}}\Big).
\end{align}
Here $c_{ij}$ with $i,j=1,2$, are normalized amplitude coefficients,
and they relate to joint probabilities in an obvious way: $
P_{ij}(\alpha ,\beta) =|c_{ij}|^{2}P_{i}(\alpha)$.

Now that the probability $P_{i}(\alpha)$ can be measured easily, as
is shown above, the value of joint probability
$P_{ij}(\alpha,\beta)$ can be determined by measuring only the
coefficients $|c_{ij}|^{2}$. This can be realized with the help of
the auxiliary photon pair $|\psi\ra_{a\bar{a}}$, which is generated
by source $S_a$ to have exactly the same form as the state under
test, i.e.,
\begin{equation}
|\psi \ra _{a\bar{a}}={\kappa }_{1}|{u}_{1}\ra _{a}|f_{1}\ra
_{\bar{a}}+{\kappa }_{2}|{u}_{2}\ra _{a}|f_{2}\ra _{\bar{a}},
\end{equation}
with the discretely entangled photon in mode $a$ heading SW and the
continuously entangled photon in mode $\bar{a}$ heading NE,
illustrated by the blue paths in Fig.~\ref{scheme}.

The auxiliary photon pair allows us to perform an indirect
measurement in the continuous-variable space of the photon in mode
$\bar{t}$. First, the mode $a$ photon of the auxiliary pair is
projected (by a polarizer $\hat{u}_1^{s}$) onto the polarization
basis $|{u}_{1}^{s}\ra$, where angle $s$ is chosen to strip off the
$|f_{2}^{\beta }\ra$ component from the photon in mode $\bar{a}$. A
glance at (\ref{stripdemo}) shows how a stripping in continuum space
by action in polarization space works. In \eqref{stripdemo}, by
choosing $\alpha$ such that $\kappa_1\tan\beta =
\kappa_2\tan\alpha$, the $|f_{2}^{\beta }\ra$ component would be eliminated. In
the case of auxiliary photon $a$, we choose $s$ such that ${\kappa
}_{1}\tan \beta = {\kappa}_{2}\tan s$ and obtain
\begin{equation}
|{u}_{1}^{s}\ra _{a~a}\la {u}_{1}^{s}|\psi \ra _{a
\bar{a}}=\sqrt{P_{1}(s)}|{u}_{1}^{s}\ra _{A}|f_{1}^{\beta }\ra
_{\bar{a}},
\end{equation}
with $P_{1}(s)=$ $_{a\bar{a}}\la \psi |{u}_{1}^{s}\ra _{a~a}\la
{u}_{1}^{s}|\psi \ra _{a\bar{a}}$. $P_i(s)$ is determined
experimentally in exactly the same way as $P_i(\alpha)$. The photon
enters mode $A$ from mode $a$ after passing the stripping polarizer
$\hat{u}_1^{s}$, as shown in Fig.~\ref{scheme}. Then the four-photon
state after the two polarization projections in modes $t$ and $a$ is
given by
\begin{eqnarray}
|\psi\ra _{T\bar{t}A\bar{a}}&=&\sqrt{P_{1}(\alpha
)P_{1}(s)}|{u}_{1}^{\alpha }\ra _{T}|{u}_{1}^{s}\ra
_{A}  \notag \\
&& \otimes \Big(c_{11}|f_{1}^{\beta }\ra
_{\bar{t}}+c_{12}|f_{2}^{\beta }\ra _{\bar{t}} \Big)\otimes
|f_{1}^{\beta }\ra _{\bar{a}}. \label{before BS}
\end{eqnarray}

Next, as shown in Fig.~\ref{scheme}, the mode $\bar{t}$ photon is
combined with the mode $\bar{a}$ photon (which is in the continuous
variable state $|f_{1}^{\beta }\ra$) by a 50:50 beam splitter (BS).
The outcome modes are denoted as $\bar{T}$ (NE) and $\bar{A}$ (SE).
The effect of the BS can be expressed as
\begin{eqnarray}
|f_{j}^{\beta }\ra _{\bar{t}} &=&\Big( |f_{j}^{\beta }\ra
_{\bar{A}}+i|f_{j}^{\beta }\ra _{\bar{T}
}\Big) /\sqrt{2}, \\
|f_{j}^{\beta }\ra _{\bar{a}} &=&\Big( i|f_{j}^{\beta }\ra
_{\bar{A}}+|f_{j}^{\beta }\ra _{\bar{T}}\Big) /\sqrt{2}.
\end{eqnarray}
As a result of Hong-Ou-Mandel (HOM) interference \cite{HOM87}, the
coincidence of the outcome photons in modes $\bar{T}$ and $\bar{A}$
determines the degree of distinguishability between the photons in
modes $\bar{t}$ and $\bar{a}$. To be more specific, the contributing
component of the mode $\bar{t}$ photon in Eq.~(\ref{before BS}) to
the coincidences after the BS is $c_{12}|f_{2}^{\beta
}\ra_{\bar{t}}$, which is the distinguishable component of the
photon in mode $\bar{a}$. This amounts to a filtering or projecting
operation of the photon in mode $\bar{t}$ onto the continuous
variable basis $|f_{2}^{\beta }\ra$.

With the above operations, a joint projection is realized for
testing the entangled photon pair $|\psi\ra_{t\bar{t}}$. It is then
straightforward to achieve the joint probability $P_{12}(\alpha
,\beta )$. The four-photon coincidence probability in modes
$T,\bar{T},A,\bar{A}$ is given as
\begin{equation}
P_{T\bar{T}A\bar{A}}(\alpha ,\beta )=\frac{P_{1}(\alpha
)P_{1}(s)|c_{12}|^{2}}{2}=\frac{N_{\alpha\beta}(T,A,\bar{T},\bar{A})}{N(t,a,\bar{t},\bar{a})},\label{4photon
coincidence}
\end{equation}
where $N_{\alpha\beta}(T,A,\bar{T},\bar{A})$ and
$N(t,a,\bar{t},\bar{a})$ are four-photon coincidence counts of the
corresponding modes for polarization angles $\alpha$ and $\beta$.
The individual probabilities can be determined using
\eqref{individual}. Consequently, the joint probability can be
written in terms of measurable quantities,
\begin{equation}
P_{12}(\alpha ,\beta
)=\frac{2N(a,\bar{a})N_{\alpha\beta}(T,A,\bar{T},\bar{A})}{N_{s}(A,\bar{a})N(t,a,\bar{t},\bar{a})}.
\end{equation}

Measurement of the other joint probabilities $P_{11}(\alpha,\beta)$
and $P_{2j}(\alpha,\beta)$ are accomplished by appropriately
rotating the angles $\alpha$ and $\beta$ by $\pi/2$. In this way the
correlation function $C(\alpha ,\beta )$ can be achieved
straightforwardly. Other correlations can be obtained similarly with
other choices of angles $\alpha $ and $\beta $. To achieve the Bell
violation given in (\ref{violate}) the orientation of the stripping
polarizer is determined as $\tan s =
(\kappa_{1}/\kappa_{2})(\sqrt{2}-1)$ and $\tan s' =
(\kappa_{1}/\kappa_{2})(\sqrt{2}+1)$.

Beyond the Bell violation issue, it is important to note that our
method of measuring the continuous-variable space is an example of
non-local quantum control \cite{nonlocal-control}. It provides a new
perspective on indirect measurement of a system state which is not
directly accessible experimentally. We have shown explicitly how, by
manipulating a discrete and controllable entangled partner,
measurements of a continuum system may be made. Apart from increased
measurement capabilities, this type of indirect measurement may be
useful for transferring or encoding information into
continuous-variable spaces which are difficult to detect or probe
directly. Therefore, with proper design, it may be possible to
construct communication protocols which impede potential
eavesdroppers from obtaining the encoded information.

In summary, we have addressed the two obstacles mentioned in
paragraph 3, obtaining a resolution with the aid of a new approach
to continuous-variable measurement. Specifically, we have devised a
Bell-CHSH inequality for the two-particle case in which one particle
is defined by an unbounded continuous variable in a unknown state of
arbitrary complexity, and we have sketched a currently feasible
measurement approach for its implementation. This technique may
expand further the systems in which Bell non-locality may be used
for practical applications \cite{Brunner2013}.

We acknowledge helpful discussions with W.P. Grice, and acknowledge
partial financial support from the following agencies: DARPA
HR0011-09-1-0008, ARO W911NF-09-1-0385, NSF PHY-0855701, and NSF
PHY-1203931.

\newpage
\newpage

\section*{Supplemental Material}

\begin{figure}[h]
\includegraphics[width=4cm]{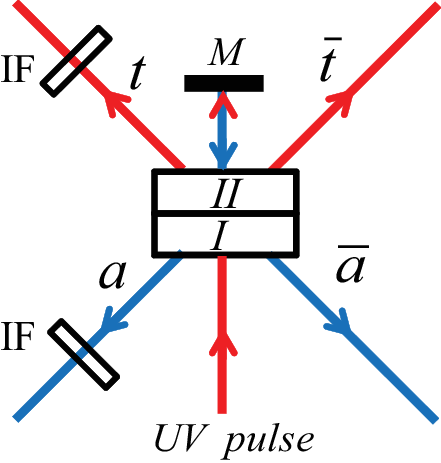}
\caption{Schematic illustration of producing two identical
discrete-continuum entangled photon pairs with two-passage
spontaneous parametric down conversion. The ultra-short UV laser
pulse passes through the combination of type I and type II crystals
and creates the first photon pair $|\psi'\ra _{t\bar{t}}$ (in red
paths) with the signal and idler photons propagating in modes $t$
and $\bar{t}$ respectively. The UV pulse pulse is then reflected
back by the mirror $M$ and passes through the two-crystal structure
again to create the second entangled photon pair $|\psi'\ra
_{a\bar{a}}$ (in blue paths) with the signal and idler photons
propagating in modes $a$ and $\bar{a}$ respectively. The spectrum of
the signal photons in modes $t$ and $a$ are filtered by the
interference filters (IF) so that the desired discrete-continuum
(polarization-spectrum) entangled states $|\psi\ra _{t\bar{t}}$ and
$|\psi\ra _{a\bar{a}}$ are achieved respectively. } \label{SPDCI&II}
\end{figure}

At least two approaches are open for generating a discrete-continuum
(e.g., polarization-spectrum) entangled state to perform the Bell
test proposed in the text. One setup is illustrated in
Fig.~\ref{SPDCI&II}, where the entangled photon pairs are produced
in a pair of spontaneous parametric down-conversion (SPDC) crystals,
a combination of type I and type II, pumped with an ultra-short UV
laser pulse in a double pass configuration \cite{Teleportation98}.
The photon pair produced by the first passage can be written in
general as
\begin{eqnarray}
|\psi' \ra _{t\bar{t}} &=&\int \int d\omega _{1}d\omega _{2}\Phi
(\omega _{1},\omega _{2})|\omega _{1},H\ra _{t}|\omega _{2},H\ra
_{\bar{t}}
\notag \\
&&+\int\int d\omega _{1}d\omega _{2}\Psi (\omega _{1},\omega
_{2})|\omega _{1},V\ra _{t}|\omega _{2},H\ra _{\bar{t}},
\label{typeI&II}
\end{eqnarray}
where $\Phi (\omega _{1},\omega _{2})$, $\Psi (\omega _{1},\omega
_{2})$ are two different amplitude functions relating to the
field-crystal interaction parameters, and $\omega _{1}$, $\omega
_{2}$ represent the frequency of the signal (mode $t$) and idler
(mode $\bar{t}$) photons respectively. Here $|\omega _{1},H\ra $
represents a single photon state with frequency $ \omega _{1}$ and
polarization $H$. The first and second terms in Eq.~(\ref{typeI&II})
are generated by the type I and type II crystals, respectively. The
propagation directions of the two down-converted photons are
determined by the phase-matching conditions of the SPDC crystals.

Then one can insert an interference filter (IF) centered at $\omega
_{0}$ in front of the signal photon (as shown in
Fig.~\ref{SPDCI&II}). After the filter the two-photon state is left
in a desired discrete-continuum (polarization-spectrum) entangled
state, i.e.,
\begin{equation}
|\psi \rangle _{t\bar{t}}=\int d\omega _{2}[\Phi _{\omega
_{0}}(\omega _{2})|H\rangle _{t}|\omega _{2}\rangle _{\bar{t}}+\Psi
_{\omega _{0}}(\omega _{2})|V\rangle _{t}|\omega _{2}\rangle
_{\bar{t}}], \label{polarzation-spectral}
\end{equation}
where $\Phi _{\omega _{0}}(\omega _{2})=\int d\omega _{1}f^{\ast
}(\omega _{1}-\omega _{0})\Phi (\omega _{1},\omega _{2})$, $\Psi
_{\omega _{0}}(\omega _{2})=\int d\omega _{1}f^{\ast }(\omega
_{1}-\omega _{0})\Psi (\omega _{1},\omega _{2})$, and $ f(\omega
_{1}-\omega _{0})$ is the spectral response function of the filter.
Here we have omitted the factorable components, i.e., the spectral
state of the photon in mode $t$ and the polarization state of photon
in mode $\bar{t}$.

While this scheme is capable in principle of generating the required
discrete-continuous entanglement, it is likely that the spontaneous
parametric down-conversion sources will have to be specially
engineered to achieve a large degree of entanglement. This is
because the degree of entanglement of the state in (2) is directly
related to the degree of orthogonality of the conditional wave
functions $\Phi_{\omega_0}(\omega_2)$, and
$\Psi_{\omega_0}(\omega_2)$. When
$\int\,d\omega_2\Phi^*_{\omega_0}(\omega_2)\Psi_{\omega_0}(\omega_2)\simeq0$,
the degree of entanglement will approach the maximal value possible.
In practice, engineering the sources to achieve this may be
difficult since the output of both crystals will have very similar
biphoton wave functions, differing only in the crystal
phase-matching functions. Regardless, the extreme control over the
biphoton wave function in spontaneous parametric down-conversion
which has been demonstrated in previous studies gives some optimism
that this obstacle may be overcome \cite{Nasr2008,Sensarn2010}.

After the first passage, the UV laser pulse is reflected back by a
mirror (M) and then passes through the two-crystal structure again
to create the second desired discrete-continuum entangled photon
pair $|\psi \ra _{a\bar{a}}$. As shown in Fig.~\ref{SPDCI&II}, the
two down-converted photons propagate in blue paths with the signal
photon in mode $a$ and the idler photon in mode $\bar{a}$. Again the
spectrum of the signal photon is filtered by an identical IF
centered at $\omega _{0}$.

Then the two photons in mode $\bar{t}$ and $\bar{a}$ can be combined
by a 50:50 beam splitter as proposed in the text to perform the Bell
test measurement. To ensure the temporal indistinguishability of the
two photons arriving at the beam splitter, one needs to make sure
that the laser pulse length is much shorter than the coherence time
of the down-converted photon \cite{Teleportation98}. By adjusting
the distance between the two-crystal structure and the mirror one
can achieve the Hong-Ou-Mandel effect \cite{HOM87}, and thus realize
the necessary temporal indistinguishability.

Another approach for realizing the necessary temporal
indistinguishability is to produce SPDC photon pairs with very long
coherence times by using a very narrow-band filter as demonstrated
in Ref.~\cite{Halder-etal2007}. The temporal indistinguishability is
then provided by appropriate post-selection of coincident detection
events in fast single-photon detectors. In this case the second
(auxiliary) discrete-continuum entangled pair can be generated from
an identical yet independent two-crystal structure.


\begin{thebibliography}{99}

\bibitem{EPR35} A. Einstein, B. Podolsky, and N. Rosen, \pr{47}, 777 (1935).

\bibitem{Bell64} J.S. Bell, Physics \textbf{1}, 195-200 (1964).

\bibitem{CHSH69} J.F. Clauser, M.A. Horne, A. Shimony and R.A. Holt,
\prl{23}, 880 (1969).

\bibitem{Clauser-Freedman} S.J. Freedman and J.F. Clauser, \prl{28}, 938
(1972).

\bibitem{Fry-Thompson} E.S. Fry and R.C. Thompson, \prl{37}, 465 (1976).

\bibitem{Aspect-etal82} A. Aspect, P. Grangier and G. Roger, \prl{49}, 91
(1982), and A. Aspect, J. Dalibard and G. Roger, \prl{49}, 1804
(1982).

\bibitem{Rowe-etal01} M.A. Rowe, D. Kielpinski, V. Meyer, C.A.
Sackett, W.M. Itano, C. Monroe, and D.J. Wineland, \nat{409}, 791
(2001).

\bibitem{Ansmann-etal09} M. Ansmann, H. Wang, R.C. Bialczak, M.
Hofheinz, E. Lucero, M. Neeley, A.D. O'Connell, D. Sank, M. Weides,
J. Wenner, A.N. Cleland, and J.M. Martinis, \nat{461}, 504 (2009).

\bibitem{Rarity1990} J.G. Rarity and P.R. Tapster, \prl{64}, 2495 (1990).

\bibitem{Kwiat1993} P.G. Kwiat, A.M. Steinberg, and R.Y. Chiao, \pra{47}, R2472-R2475 (1993).

\bibitem{Strekalov1996} D.V. Strekalov, T.B. Pittman, A.V. Sergienko, Y.H. Shih, and P.G. Kwiat, \pra{54}, R1-R4 (1996).

\bibitem{Kuzmich2000} A. Kuzmich, I.A. Walmsley, and L. Mandel, \prl{85}, 1349 (2000).

\bibitem{Yarnall-etal07} T. Yarnall, A.F. Abouraddy, B.E.A. Saleh, and
M.C. Teich, \prl{99}, 170408 (2007).

\bibitem{Franson1989} J.D. Franson, \prl{62}, 2205 (1989).

\bibitem{Banaszek} K. Banaszek and K. W\'{o}dkiewicz, \pra{58}, 4345 (1998); \prl{82}, 2009 (1999); Acta Phys. Slovaca {\bf 49}, 491 (1999).

\bibitem{Gilchrist-etal98} A. Gilchrist, P. Deuar, and M.D. Reid,
\prl{80}, 3169 (1998).

\bibitem{Munro-Milburn98} W.J. Munro and G.J. Milburn, \prl{81}, 4285
(1998); W.J. Munro, \pra{59}, 4197 (1999).

\bibitem{Chen-etal02} Z.B. Chen, J.W. Pan, G. Hou, and Y.D. Zhang,
\prl{88}, 040406 (2002).

\bibitem{Brukner2003} C. Brukner, M.S. Kim, J.-W. Pan, and A.
Zeilinger, \pra{68}, 062105 (2003).

\bibitem{Wenger-etal03} J. Wenger, M. Hafezi, F. Grosshans, R.
Tualle-Brouri, and P. Grangier, \pra{67}, 012105 (2003).

\bibitem{Broadbent-etal} C.J. Broadbent, X.-F. Qian and J.H. Eberly,
to be submitted.

\bibitem{CFRD} E.G. Cavalcanti, C.J. Foster, M.D.
Reid, and P.D. Drummond, \prl{99}, 210405 (2007). see Q. Sun, H. Nha, and M.S. Zubairy, Phys. Rev. A 80,
020101(R), (2009). see A. Salles and D.
Cavalcanti and A. Ac\'{i}n, \prl{101}, 040404 (2008).


\bibitem{Brunner2013} These technologies include quantum-assisted
communication complexity, quantum-assisted zero-error communication,
device-independent quantum key distribution, and device-independent
randomness generation. For a brief introduction to these topics see
N. Brunner, D. Cavalcanti, S. Pironio, V. Scarani, and S. Wehner,
arXiv:1303.2849, p. 28-35.


\bibitem{Schmidt} E. Schmidt, Math. Ann. {\bf 63}, 433 (1907). See also
A. Ekert and P.L. Knight, \ajp{63}, 415 (1995) and J.H. Eberly,
Laser Phys. {\bf 16}, 921 (2006).

\bibitem{Qian-Eberly11} A similar but fully classical situation has been
examined in X.-F. Qian and J.H. Eberly, \ol{36}, 4110 (2011).


\bibitem{Gisin} N. Gisin, \pla{154}, 201 (1991).


\bibitem{HOM87} C.K. Hong, Z.Y. Ou, and L. Mandel, \prl{59}, 2044 (1987).

\bibitem{nonlocal-control} See, for example, J. Eisert, K. Jacobs, P.
Papadopoulos, and M.B. Plenio, \pra{62}, 052317 (2000); D. Collins,
N. Linden, and S. Popescu, \pra{64}, 032302 (2001). In fact, the
setup in Fig.~\ref{scheme} is essentially the same as that used by
M. Pavi\v{c}i\'{c} and J. Summhammer, \prl{73}, 3191 (1994), in an
early entanglement swapping experiment.

\end{thebibliography}

\begin{thebibliography}{99}

\bibitem{Teleportation98} D. Bouwmeester, J.-W. Pan, K. Mattle, M. Eibl,
H. Weinfurter, and A. Zeilinger, \nat{390}, 575 (1997).


\bibitem{Nasr2008} M.B. Nasr, S. Carrasco, B.E.A. Saleh, A.V. Sergienko, M.C. Teich, J.P.
Torres, L. Torner, D.S. Hum, and M.M. Fejer, \prl{100}, 183601
(2008).

\bibitem{Sensarn2010} S. Sensarn, G.Y. Yin, and S.E. Harris, \prl{104}, 253602 (2010).

\bibitem{HOM87} C. K. Hong, Z. Y. Ou, and L. Mandel, \prl{59}, 2044 (1987).

\bibitem{Halder-etal2007} M. Halder, A. Beveratos, N. Gisin, V. Scarant, C.
Simon, and H. Zbinden, \nphys{3}, 692 (2007).


\end{thebibliography}
\end{document}